\documentclass[aps,prb,twocolumn,superscriptaddress,showpacs]{revtex4}
\usepackage{amsmath}
\usepackage{bm}
\usepackage{subfigure}
\usepackage[dvipdfmx]{graphicx}

\begin{document}

\title{Half-integer contributions to the quantum Hall conductivity from single Dirac cones}
\author{Haruki Watanabe}
\email[]{watanabe@cms.phys.s.u-tokyo.ac.jp}
\affiliation{Department of Physics, University of Tokyo, Hongo, Tokyo 113-0033, Japan}
\author{Yasuhiro Hatsugai}
\affiliation{Institute of Physics, University of Tsukuba, Tsukuba, 305-8571, Japan}
\author{Hideo Aoki}
\affiliation{Department of Physics, University of Tokyo, Hongo, Tokyo 113-0033, Japan}
\date{July 31, 2010}

\begin{abstract}
While the quantum Hall effect in graphene has been regarded as a realization of the {\em anomaly} associated with the massless Dirac particle carrying half the usual topological integer, this is hidden due to the doubling of the Dirac cones. 
In order to confirm the half-integer contribution from each Dirac cone, here we theoretically consider a lattice model in which the relative energy between the two Dirac points is systematically shifted. 
With an explicit calculation of the topological (Chern) number, we have demonstrated that each Dirac cone does indeed contribute to the Hall conductivity as the half odd integer series ($\cdots, -3/2, -1/2, 1/2, 3/2, \cdots$) when the Fermi energy traverses the (shifted sets of) Landau levels. 
The picture is also endorsed, via the bulk-edge correspondence, from the edge mode spectrum for the present model. 
\end{abstract}


\maketitle

\section{Introduction} 
In the seminal discovery of the quantum Hall effect (QHE) in graphene\cite{Nov05, Kim05, NetoDirac}, a most striking point is that the graphene QHE is regarded as a realization of the {\em anomaly}\cite{DNR} in the massless Dirac particle, where each Dirac cone carries 1/2 of the usual QHE topological number\cite{TKNN, HatsugaiBulkEdge93}.
Namely, in place of the usual Hall conductivity, $\sigma_{xy} = 0, 1, 2, \cdots$ (in units of $-e^2/h$, with the spin degrees of freedom dropped here), we have $\sigma_{xy} = 2n+1 (n = 0, \pm1, \cdots)$. 
The standard understanding is that the honeycomb lattice has a couple of Dirac cones at K and K' points in the Brillouin zone, so that we just divide the above formula by 2 to have $\sigma_{xy} = n+1/2$ for the contribution from each valley. 
The half integers are thus hidden in the total Hall conductivity. 
Since the Nielsen-Ninomiya theorem\cite{Ninomiya} dictates that Dirac cones should always appear in pairs as far as the chiral symmetry is present, we can pose an important question: is it really impossible to resolve the half-integer components? 

In terms of field theory (as opposed to lattice models), the situation is simple: the Hall conductivity for a massive ($m\neq 0$) Dirac particle, in zero magnetic field, is given by $\sigma_{xy}=(1/2)\mathrm{sgn}(m)$ when the Fermi energy $E_F$ is in the mass gap\cite{DNR}. This can be readily shown by calculating Berry's connection. 
If we consider a field theoretical model concerning a {\it single} Dirac fermion at $\bm{k}_0$ with a gap $|m|$, the Hamiltonian is given by $h=\delta k_x\sigma_x+\delta k_y\sigma_y+m\sigma _z=\bm{R}\cdot\bm{\sigma}$, where $\bm{R}=(R_1,R_2,R_3)=(\delta k_x,\delta k_y ,m)$, $\delta\bm{k}=\bm{k}-\bm{k}_0$\cite{Berry}. 
If we denote the eigenstate having an energy $-R=-\sqrt{|\delta\bm{k}|^2+m^2}$ as $\psi$, and its Berry connection as ${\cal A}=\psi^\dagger d\psi$, the total Berry curvature over the $(k_x,k_y)$ plane (represented as $\Pi_m\equiv \{\bm{R}\big|R_3=m\}$) is $C=\frac{1}{2\pi i}\int _{\Pi_m}d{\cal A}=\int _{\Pi_m}\bm{B}\cdot d\bm{S}={\rm sgn}\,(m)/2$, where $\bm{B}=\bm{R}/(4\pi R^3)$, which is the flux from a magnetic monopole at the origin with ${\rm div}\,\bm{B}=\delta^3(\bm{R})$ and $d\bm{S}= dk_xdk_y\, \partial_{k_x}\bm{R}\times\partial_{k_y}\bm{R}=dk_xdk_y(0,0,1)$. 
We can then see that we have 1/2 since just the half the total flux of the magnetic monopole passes through the plane $\Pi_m$. 
This implies that the topological change in the quantum ground state (filled Dirac sea) is characterized by $\Delta \sigma_{xy}=\pm 1$\cite{Haldane, Oshikawa94, Hatsugai96}.
Generically speaking, a topological quantum phase transition is naturally accompanied by a sign change in the mass in the effective, low-energy Dirac fermions. 
Quantum Hall plateau transition is a typical example, and the topological insulators such as the quantum spin Hall system are also described along this line, where the spin-orbit interaction induces the sign change in the mass\cite{Konig07, KaneMele, HasanKane}. 

While the quantization into the half odd integers is conceptually interesting, if we go over to lattice models, however, we have a periodicity in the Brillouin zone, which implies that the topological numbers should always be integers as dictated by the celebrated but inescapable Thouless-Kohmoto-Nightingale-den Nijs (TKNN) formula\cite{TKNN}.
The integer total Hall conductivity for graphene with a pair of Dirac cones is an example of this. 
If we turn to a wider class of lattice models, we can go around the Nielsen-Ninomiya theorem.
For instance, we can have a lattice model that has an odd number of massless Dirac cones, but even in that case we still end up with integer Hall conductivities, which may be regarded as due to hidden massive Dirac fermions required to exist for the topological consistency in the lattice system to be guaranteed. 
The massive Dirac fermions (sometimes regarded as ``spectators"\cite{Hatsugai96}) are hidden in the high-energy region, which do not appear in the low-energy physics except that they make sure that the the Hall conductivity are topologically protected to be integers. 
In another manipulation of Dirac cones\cite{Haldane}, a quantum Hall effect in zero total magnetic field has been considered, where the Hall conductivity is shown to take the value of $e^2/h$ even in zero field in a model containing complex hoppings in the situation where the Dirac cones are made massive. 
It is thus rather difficult to confirm half-odd-integers in the Hall conductivity for a single Dirac cone. 

In this Rapid Communication, we want to shed light to this problem by posing the following question: if we can manipulate the energies of the multiple Dirac points, can the half integer series confirmed through a systematic behavior of the total Hall conductivity?  
Namely, we shall construct a lattice model, where the two Dirac points are shifted with the massless cones preserved for both of the Dirac points. 
We shall identify, by directly computing the topological Chern number for systematically varied relative position of the Dirac cones, 
that each Dirac cone indeed has a half-odd-integer series ($\cdots, -3/2, -1/2, 1/2, 3/2, \cdots$) when $E_F$ traverses (now shifted sets of) Landau levels belonging to the two Dirac cones. 
The picture is further endorsed, via the bulk-edge correspondence, from the behavior of the edge modes for the shifted Dirac cones for finite systems. 

\begin{figure}[!t]
\begin{minipage}[!ht]{0.39\hsize}
\subfigure[]{\includegraphics[width=80pt]{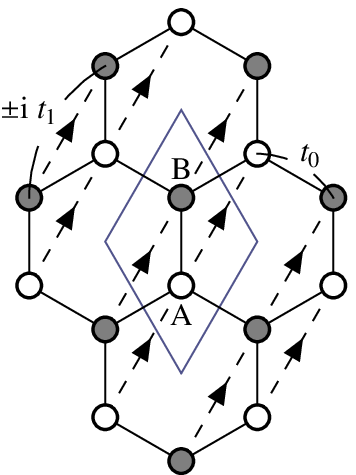}\label{fig:model}}
\end{minipage}
\begin{minipage}[!ht]{0.59\hsize}
\subfigure[]{\includegraphics[width=140pt]{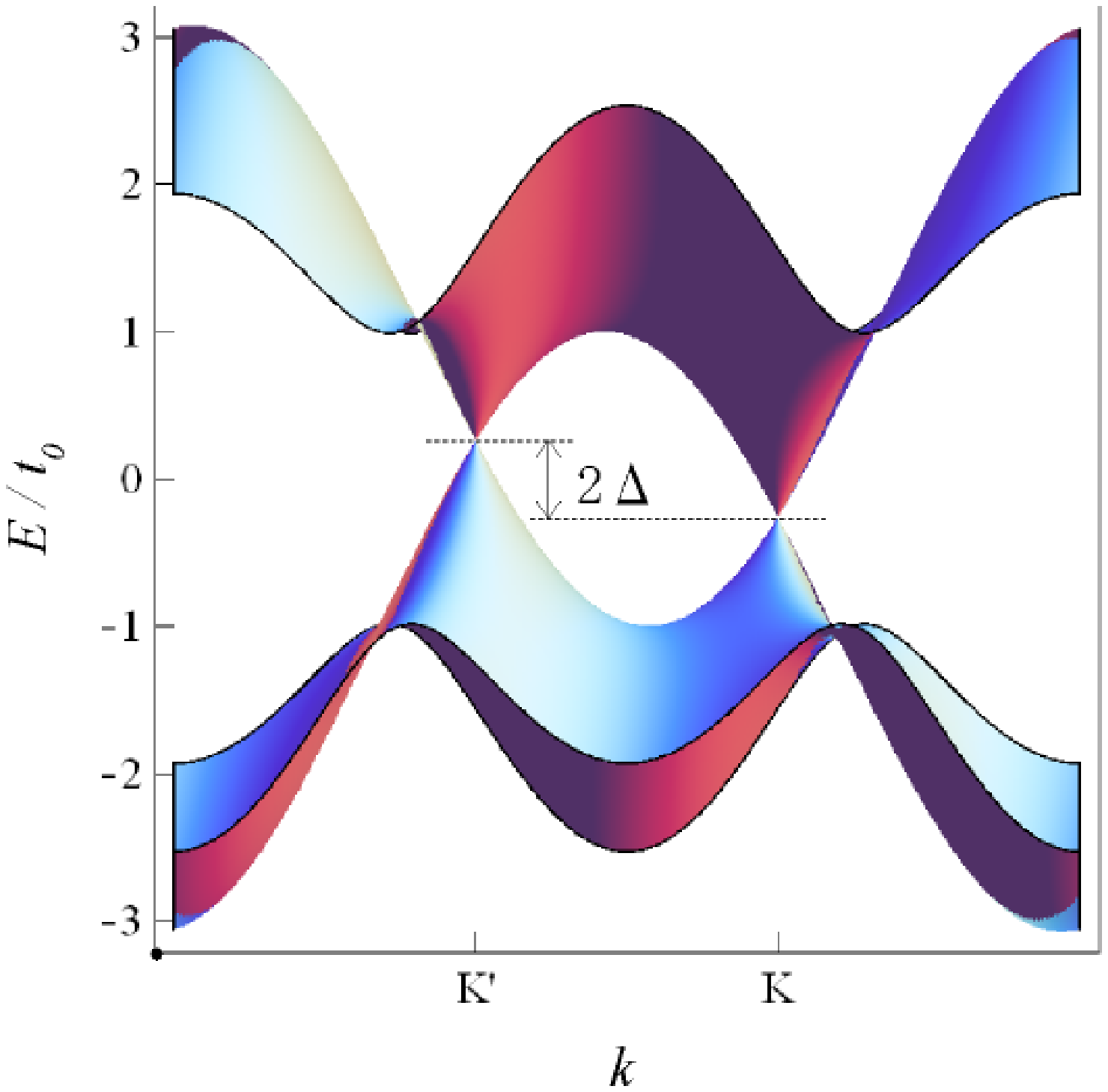}\label{fig:dirac}}
\end{minipage}
\caption{(Color online) (a)The honeycomb lattice model considered here with nearest-neighbor hopping $t_0$ (solid lines) and second-neighbor hopping which is $i t_1$ along each dashed arrow, $-it_1$ in the opposite direction. A unit cell and $A$, $B$ sublattices are indicated. 
(b)The dispersion relation in the present model, seen horizontally, with Dirac cones shifted in energy by $\Delta=\sqrt{3}t_1$.}
\label{fig:dispersion}
\end{figure}

\section{Shifted Dirac cones} 
A strategy to conceive a model in which two Dirac cones are preserved but mutually shifted in energy is the following. 
A simplest solution is to add a term that is proportional to $\sigma_0$ (unit matrix) in the space of Pauli matrices with a $\bm{k}$-dependent coefficient. 
Thus we can introduce a Hamiltonian,
\begin{gather}
\label{eq:hamiltonian}
{\cal H}=
\sum_{\bm{k}}\sum_{\alpha,\beta}\hat{c}^{\dagger}_{\bm{k}\alpha}[h^{\mathrm{gr}}_{\bm{k}}+2t_1(\sin k_1)\sigma_0]_{\alpha,\beta}\hat{c}_{\bm{k}\beta},\\
\notag
h^{\mathrm{gr}}_{\bm{k}}=t_0\big[(1+\cos{k_1}+\cos{k_2})\sigma_1+(\sin{k_1}+\sin{k_2})\sigma_2\big],
\end{gather}
where $\hat{c}^{\dagger}_{\bm{k}\alpha}$ creates an electron in $\bm{k}$- space, and $\sigma_i$'s are Pauli matrices with $\alpha, \beta$ denoting their components. 
In this Hamiltonian, we have added, on top of the nearest-neighbor hopping $t_0$, an extra $\sigma_0$ term with hopping $t_1$ as a coefficient. 
This only affects the band dispersion as an additive, $\bm{k}$-dependent term, which implies that we lift the degeneracy between K and K' points if the $\bm{k}$-dependent term has different values at K and K'. 
A simplest choice is $\propto {\rm sin}k_1$. 
If we go back to the real space, the tight-binding model is as depicted in Fig.\ref{fig:model}, which has extra second-neighbor hoppings. 
The added hopping has to be only between $A$-$A$ and $B$-$B$ for the Dirac cone to be preserved, and they have to be complex for the degeneracy between K and K' (mutually time-reversal pairs) to be lifted. 
So the model with a complex hopping is rather fictitious, but we do accomplish shifted Dirac cones as depicted in Fig.\ref{fig:dirac}. 
In this model the chiral symmetry [i.e., the presence of an operator ($\sigma_z$ in the case of graphene) that anticommutes with the Hamiltonian] is broken, since appearance of $\sigma_0$ in ${\cal H}$ invalidates $\{{\cal H}, \sigma_z\}=0$. 
Nevertheless, the addition of $\sigma_0$ preserves the shape of Dirac cones, along with the species doubling. If we expand Hamiltonian (\ref{eq:hamiltonian}) around $\bm{k}_0$ (K or K' points), we have
\begin{equation}
\label{eq:DF}
h_{\bm{k}}\simeq\chi\Delta\,\sigma_0-\hbar v_F\big[\chi\delta k_x\sigma_1+\delta k_y\sigma_2\big],
\end{equation}
where $\chi=+1(-1)$ correspond to K (K'), $\Delta=\sqrt{3}t_1$ is (half) the shift, $v_F=\sqrt{3}a\,t_0/2\hbar$ the Fermi velocity, and $\delta\bm{k} = \bm{k} - \bm{k}_0$. 
So the effective theory around these points are a pair of Dirac equations at K and K' with a shift $\pm\Delta$ in energy as desired.

\section{Chern numbers} 
The Landau levels in a magnetic field $B_z$ expected from the effective Dirac Eq. (\ref{eq:DF}) are $\varepsilon_n^{\chi}=\chi\Delta+\hbar\omega_c\mathrm{sgn}(n)\sqrt{n}$ \,[$\omega_c=3^{1/4}t_0\sqrt{2\pi \phi}$, $\phi = (\sqrt{3}a^2 B_z/2)/(hc/e), n = 0, \pm 1, \pm 2, \cdots$]. 
However, we should of course go back to the original lattice model for obtaining the Landau levels and, especially, the Hall conductivity. 
The calculation of the Hall conductivity, which is a topological Chern number, requires a bit of care, since we have to question the 
behavior of the topological number around $E=0$ which is just where the hole and electron branches meet in the Dirac cone dispersions. 
This implies we have to sum over all the contributions from the ``Dirac sea" to get the total Hall conductivity. 
This ordinarily poses a numerically difficult problem, but we can overcome the problem with a method that employs a non-commutative Berry's connection\cite{HatsugaiBerry} and its integration (Chern number) over the Brillouin zone with a technique developed in the lattice gauge theory\cite{Fukui}. 

\begin{figure}[!ht]
\begin{center}
\includegraphics[width=200pt,clip]{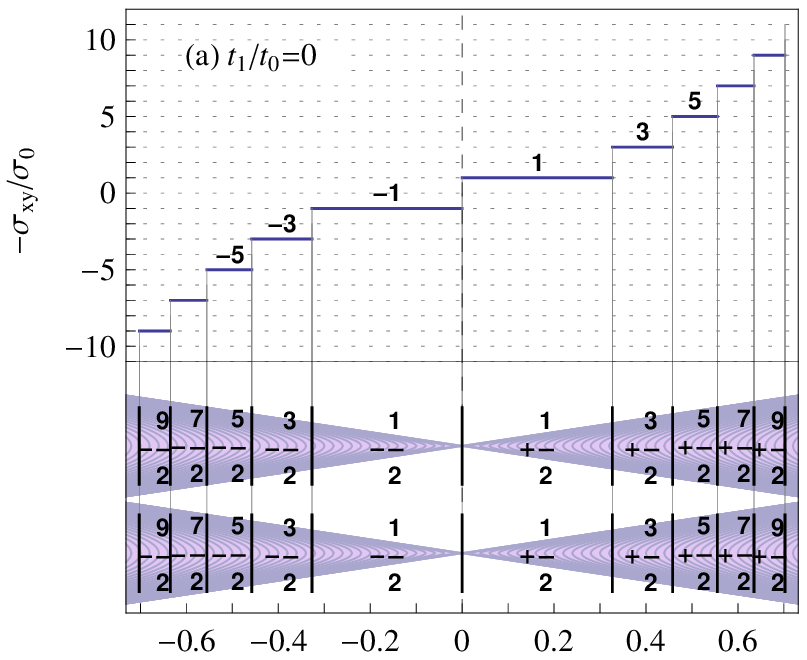}\label{fig:LLchern0}
\includegraphics[width=200pt,clip]{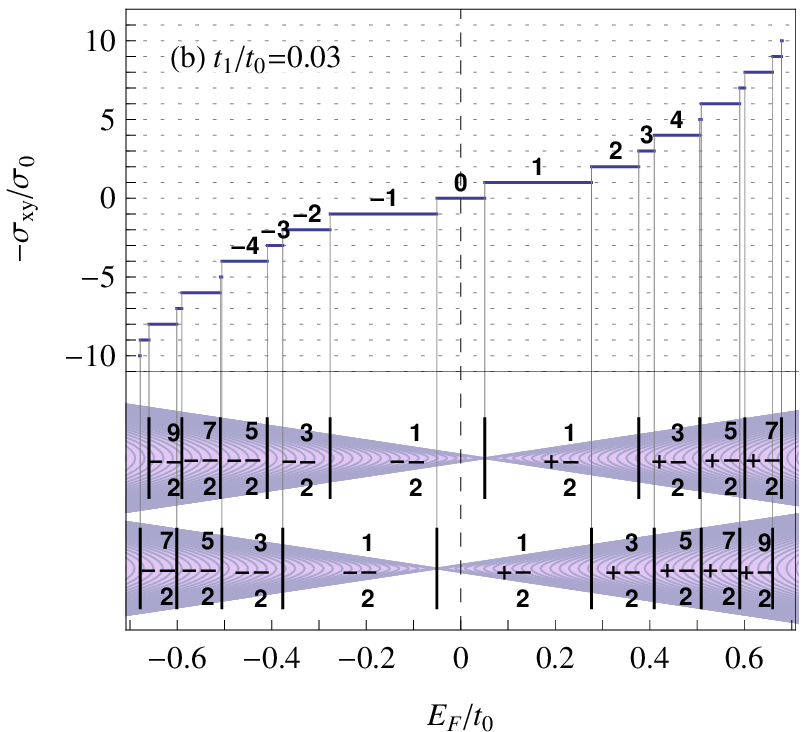}\label{fig:LLchern3}
\caption{(Color online) (a) For the ordinary graphene ($t_1=0$) and (b) for the present model with a shifted Dirac cones ($t_1 \neq 0$) we display the numerically calculated Chern numbers against $E_F$ (upper panels), along with the Landau levels for each of the two Dirac cones, where the Chern numbers are indicated for $E_F$ in each of the gaps (lower panels). 
The result is for the flux $\phi=1/100$. 
\label{fig:LLchern}}
\end{center}
\end{figure}

The result for the Chern number in the present model obtained with the above method is displayed in Fig.\ref{fig:LLchern}, which is the key result in the present work, and reveals an intriguing feature. 
For comparison, let us first look at Fig.\ref{fig:LLchern} (a), which is just the result for the ordinary graphene with the usual graphene Landau levels at $\varepsilon_n^{\chi}=\hbar\omega_c\mathrm{sgn}(n)\sqrt{n}$. 
The numerical result for the Chern number is simply the graphene QHE, where the well-known Chern number sequence of $\cdots, -3, -1, 1, 3, \cdots$ for the honeycomb lattice coincides with a sum over the two, half-integer contributions from the degenerate pair of Dirac fields. 

\begin{figure}[]
\includegraphics[width=150pt,clip]{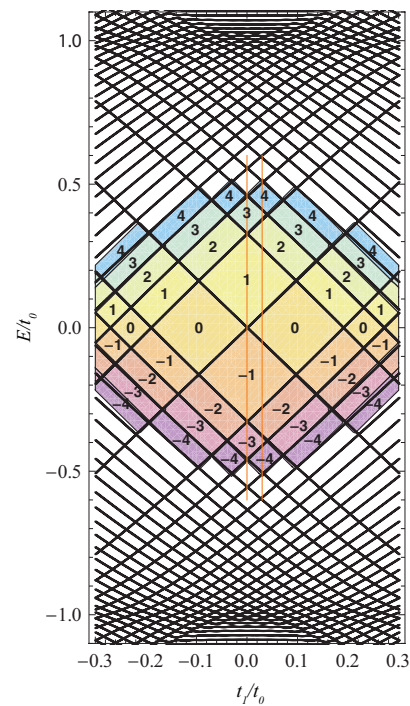}
\caption{(Color online) Numerical result (black curves) for the Landau level spectrum when the shift ($\Delta \propto t_1$; the horizontal axis) is varied, with the Chern number indicated for each of the Landau gap. 
The shading represents the Chern numbers expected from the effective theory (see text), where the boundaries between different shades (Chern numbers) almost exactly coincide with the numerical result. 
Accumulation of lines around $E=\pm 1$ corresponds to the Van Hove singularity, which shifts quadratically with $t_1$. 
The result is for a magnetic field of $\phi=1/100$, and two vertical lines indicate, respectively, the situation for Figs.\ref{fig:LLchern}(a) and \ref{fig:LLchern}(b). 
\label{fig:landaulevel}}
\end{figure}

If we now turn to the result for the present lattice model in Fig.\ref{fig:LLchern} (b), the crucial question is: does the result conform to the expectation from the effective Dirac model where the sum of two half-odd-integer series ($\cdots, -3/2, -1/2, 1/2, 3/2, \cdots$) with a shift in energy between them should give the total Hall conductivity as $E_F$ traverses shifted sets of Landau levels (as displayed in the lower panel in the figure). 
The numerical result for the Chern number in the lattice model is seen to agree with this. Namely, in a striking contrast to the ordinary graphene where each QHE step has a jump of 2 in the Chern number, the present model exhibits a jump of 1 at each step. 
This, along with the positions of these jumps, perfectly fit with the positions and the associated Chern numbers of the shifted Landau levels as the figure indicates. 
The agreement is rather surprising, since there is no obvious reason why the superposition of effective field theory for the vicinities of K and K' and the lattice model should have the same Chern numbers. 
Thus, although we have still no half integers for the total Hall conductivity (since a sum of two half-odd integers is an integer), we have indirectly confirmed the half integers. 
Since no half integers can possibly appear according to TKNN, this is indeed {\it as best as} we can confirm the half-integer property of each Dirac cone. 

We expect such a decomposition of the Hall conductivity into the contributions from each Dirac cone to systematically occur when we vary the energy separation ($2\Delta$) between the two Dirac points. 
Fig.\ref{fig:landaulevel} displays the loci of Landau levels (solid lines) and the Chern number for each gap between the Landau levels when $t_1 \propto \Delta$ is varied. 
The result is just what is expected from the above decomposition, as indicated by the result for the two Dirac fields (shaded regions), which almost exactly coincides with the result for the lattice model. 

\section{Edge states}
\begin{figure}[!t]
\begin{center}
\vspace{2.5\baselineskip}
\includegraphics[width=200pt]{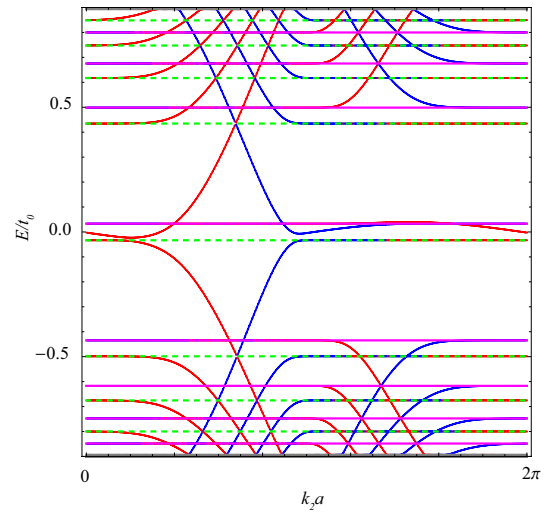}
\caption{(Color online) Energy spectrum against $k_2$ in the present model for a finite sample with zigzag-bearded edges
\cite{HFA, HatsugaiEdge93}. 
Horizontal solid (dashed) lines correspond to bulk Landau levels corresponding to the shifted Dirac fermions by $\Delta$ $(-\Delta)$. 
Curves are the edge modes whose amplitudes are confirmed to be localized along the zigzag right (red) or bearded left (blue) edge. 
The result is for a magnetic field $\phi=1/48$ with $t_1/t_0=0.02$.}
\label{fig:edge}
\end{center}
\end{figure}

The QHE state is a topological state, which immediately implies that edge states should appear at the boundaries of a sample, where an intimate relation exists between the edge topological number (i.e., the edge Hall conductivity) and the bulk topological number (the bulk Hall conductivity)\cite{HatsugaiEdge93, HatsugaiBulkEdge93}. 
So it is an intriguing as well as necessary test to look into how this property appears in the edge-state in the present model. 
Fig.\ref{fig:edge} displays a numerical result for the energy spectrum against $k_2$ for a finite sample with zig-zag edges along $x_1$. 
The spectrum comprises the bulk Landau levels (horizontal lines) and a series of edge modes (identified as curves that traverse between Landau levels). 

According to the bulk-edge correspondence in a topological argument \cite{HatsugaiBulkEdge93}, the bulk Chern number should coincide with the number of pairs of (right and left) edge modes in the Landau gap considered. 
In Fig.\ref{fig:edge} we can clearly recognize that the bulk Dirac Landau levels ($\propto\sqrt{n}$ with the shift) (horizontal lines) are accompanied by right and left edge modes (upturning for the electron branch, down-turning for the hole branch).  
This occurs around each of K and K'. 
If we count the number of pairs of edge modes, we have, for each of the series, the beautiful sequence $1, 2, 3, \cdots$ (as opposed to the sequence $1, 3, 5, \cdots$ for graphene, see Fig.10 of \cite{HFA}) both in the electron or hole branches. 
The two series of pairs of edge modes arise with the energy shift between K and K' as they should. 
We can thus reinforce our argument, via the bulk-edge correspondence, that we can decompose the contribution from each of the Dirac cone that carries the bulk Chern number of 1/2 (Fig.\ref{fig:LLchern}). 

\section{Summary} 
We have demonstrated that each Dirac fermion contributes the half odd integer series ($\cdots, -3/2, -1/2, 1/2, 3/2, \cdots$) to the Hall conductivity when the Fermi energy traverses the Landau levels in a lattice model in which the energy between the two Dirac points is systematically shifted.
The picture is also confirmed, via the bulk-edge correspondence, in the edge mode of the Dirac fermions. 

Recently it is recognized that massless Dirac fermions can be realized as surface states of the three-dimensional topological (quantum spin Hall) systems\cite{HasanKane}. 
There, the two-dimensional surface of the gapped three-dimensional topological insulator accommodates gapless (massless Dirac) fermions, which is a manifestation of the bulk-edge correspondence\cite{HatsugaiBulkEdge93}. 
The massless Dirac fermions are again doubled for the topological consistency, but the doubling partner exists at the other side of the system in this case. 
Thus, whether we can decompose the topological numbers into contributions from each Dirac cone is becoming a realistic question. 
It is an interesting future problem to consider how the present toy model can be realized in actual materials, where the surfaces of 3D topological insulators may be one possible avenue. 
Another possible realization of our model is a cleverly arranged cold atoms for an optical lattice where non-Abelian gauge structure can arise from the intra-atomic degree of freedom with a fine-tuning of desired parameters\cite{goldman}.

The work was supported in part by Grants-in-Aid for Scientific Research No. 20340098 (YH and HA) from JSPS and No. 22014002(YH) on Priority Areas from MEXT.

\end{document}